# Estimating the coronal and chromospheric magnetic fields of solar active regions as observed with the Nobeyama Radioheliograph Compared with the Extrapolated Linear Force-Free Field

A. Mouner, Abdelrazek M. K. Shaltout, M. M. Beheary, K.A.K. Gadallah, K. A. Edris

Department of Astronomy and Meteorology, Faculty of Science, Al-Azhar University, Nasr City, Cairo 11884, Egypt

**Abstract**: Adopting the thermal bremsstrahlung or so-called free-free emission process, the coronal and chromospheric magnetic fields are derived from the polarization and spectral observations with the Nobeyama Radioheliograph (NoRH) at 17 GHz. The solar active regions (ARs) located near the disk center observed on January 8, 2015 (NOAA 12257) and December 4, 2016 (NOAA 12615) are used for the estimate of the chromospheric and coronal magnetic fields with the microwave radio observations. We compare solar radio maps of active regions for both intensity and circularly polarized component with the photospheric magnetograms from observations with the Helioseismic & Magnetic Imager (HMI) and the chromosphere-corona transition region images obtained with the Atmospheric Imaging Assembly (AIA), on board the Solar Dynamic Observatory (SDO). From our analysis, we find a different structure in the radio intensity maps between two active regions owing to possibly the differential rotation of the Sun, where the AR 12257 clearly shows a widespread structure of radio intensity, but in the case of AR 12615 which exhibits a narrow structure in the total intensity map. We notice from the comparison between radio maps of both ARs that the circular polarization degree in the AR 12257 is about 2 %, but the AR 12615 has a higher existent value by 3 %. Radio observations provide us for direct measurements of magnetic fields in the chromospheric and coronal layers. We estimate the coronal magnetic fields using the AIA observations by adopting magnetic loops in the corona over some patches with weak photospheric magnetic fields. However, the coronal magnetic field derived from the SDO/AIA data was 90 - 240 G. We also study the coronal magnetic fields based on the structure of the extrapolated field, where the result of the magnetic fields was in the range 35 – 145 G, showing that the difference in the coronal magnetic fields between both results is attributed to the assumption of the force-free approximation.

## 1. Introduction

The measurements of the magnetic field of the solar atmosphere is an important key for understanding many solar phenomena such as solar flares, coronal mass ejection or solar eruptions, coronal heating and sunspots. The coronal magnetic fields can be estimated from different methods based on the structure of the extrapolated field above the photospheric boundary using a nonlinear force-free field approximation (Moraitis et al. 2016). Inoue et al. (2012) studied the structure of the extrapolated magnetic fields by adopting a linear or nonlinear force-free approximations. However, the longitudinal magnetic field of the extrapolations of potential field was estimated in some investigations by (Sakurai 1982; Shiota et al. 2008; Miyawaki et al. 2016).

Based on radio observations, the longitudinal magnetic fields can be calculated from the thermal free-free emission process in some analyses during the past decades (Bogod and Gelfreikh 1980; Iwai and Shibasaki 2013; Miyawaki et al. 2016). More recently, Miyawaki et al. (2016) computed the coronal magnetic field from polarization observations of solar active region using the NoRH data at 17 GHz, adopting the thermal free-free emission approach. However, their method focused on determining the coronal magnetic field from selecting some regions of coronal loops and weak magnetic field strength in the chromosphere along the line of sight. In this analysis, the coronal magnetic fields were found to be in the range 100–210 G. Furthermore, they derived the coronal magnetic fields from the potential field extrapolations with the values obtained from EUV observations, the estimated magnetic field strengths were 48-90 G.

In this paper, the coronal and chromospheric magnetic fields are evaluated from the radio polarization and spectral observations of the thermal free-free emission with the NoRH data at 17 and 34 GHz. The radio magnetic fields of active regions are compared with the photospheric magnetograms from observations taken by HMI and the chromosphere-corona maps with the AIA. The extrapolated field structure in the corona above the photospheric magnetograms is shown. In Section 2, more details on the NoRH and SDO observations are illustrated, while data analysis is described in Section 3. In the end, summary and discussion are presented in Section 4.

## 2. Observation

### 2.1. NoRH and SDO instruments

NoRH is a radio interferometer installed to observe the Sun to retrieve the structure of the magnetic fields in the active regions. It consists of 84 parabolic antennas with 80 cm diameter, and more details on the NoRH data is described in Nakajima et al. (1994). The spatial resolution of the radio magnetogram is about 10″ at 17 GHz and 5″ at 34 GHz. This radio interferometer is used to observe the full solar disk at 17 GHz (both in circularly polarized component and in intensity) and 34 GHz (intensity), where the radio data refer to the upper levels of the chromosphere. The solar radio images are synthesized every 10-seconds procedure. The Solar Dynamics Observatory is a NASA mission which launched on February 11, 2010. The photospheric magnetograms are observed with HMI (Scherrer et al. 2012) and the chromosphere-corona transition region are obtained with the AIA (Lemen et al. 2012) for providing EUV observations at 171 Å and 304 Å.

Data from generations of Geostationary Operational Environmental Satellites (GOES) are shown for active regions. GOES satellite carries an X-Ray Sensor (XRS) to measure the total solar flux at two wavelengths bands, 0.5 – 0.4 Å and 1 - 8 Å. In Figure 1, the time variation of the total soft X-ray flux of the Sun observed with GOES 15. The AR 12615 remained in the B-class between 22:00 UT on December 3, 2016 and 4:00 UT on December 4, 2016, while the AR 12257 stayed in the C-class between 22:00 UT on January 7, 2015 and 4:00 UT on January 8, 2015. The solar AR 12257 and 12615 are used for retrieving the distribution of the magnetic field strengths based on the theory of generation and propagation of radio emission in cosmic plasma (Alissandrakis 1999; Gelfreikh 1994; Lee et al. 1993; Iwai & Shibasaki 2013). Figure 2 shows the photospheric magnetograms (left panel) and radio intensity (right panel) maps for two ARs. There were two ARs near the disk center located at (NOAA 12257: N05W14; NOAA 12615: S07W19). The longitudinal magnetic field can be derived from the thermal bremsstrahlung process proposed by Bogod and Gelfreikh (1980), the reason for this is that the largest magnetic field in the two active regions with HMI is less than 2000 G.

**2.2. Estimates of Radio Magnetic Field**

Based on the free-free emission process, the magnetic fields in coronal and chromospheric layers are measured using single-frequency observations with the NoRH at 17 GHz. The estimate of magnetic field strength was developed by Bogod & Gelfreikh (1980). The longitudinal component of the magnetic field, $B_l$, can be defined as follows:

$$B_l[G] = \frac{10700}{n\lambda[cm]} \frac{V}{I} \quad ; \quad n = \frac{d(logI)}{d(log\lambda)} \tag{1}$$

where $B_l$ is the longitudinal component of the magnetic field, V is the brightness temperature of the circularly polarized component used to find the longitudinal component of the field, I is the brightness temperature, where radio intensity includes emissions in both coronal loop plasma and chromosphere, λ is the wavelength of observations (1.76 cm), and n is the power-law spectral index of the brightness temperature between 17 GHz and 34 GHz.

## 3. Data analysis

### 3.1. Data average time and noise level

The observation time was taken around 3:00 UT in which the raw images are synthesized using the Koshix program with interval 10 seconds (http://solar.nro.nao.ac.jp/norh/doc/manuale/index.html). All images were considered for a better sensitivity of data. Taking radio magnetograms into account, we adopted averaging the images to reduce noise levels of the data, due to the degree of circular polarization is very small. The NoRH data give us information for both the brightness temperature (Stokes-I) and circular polarization (Stokes-V) in units of Kelvin. The observed Stokes parameters *I* and *V* are defined by:

$$I = \frac{R + L}{2} \qquad V = \frac{R - L}{2} \qquad (2)$$

where the parameters R and L are the brightness temperatures of the right- and left-handed circular polarized component, respectively. A degree of polarization (*P)* is given by:

$$P = \frac{V}{I} = \frac{R - L}{R + L} \qquad (3)$$

In this analysis, the law of propagation errors method of Iwai and Shibasaki (2013) is also adopted for the estimate of standard deviations of circularly polarized signal and degree of polarization over 4-minute intervals in the analyzed data. This method was applied for the selected radio-quiet Sun regions observed on 08-Jan-2015 and 04-Dec-2016.

**Table 1.** The standard deviation of circularly polarized signal (V), minimum detectable signal level of circular polarized (5σ) and average intensity ($I_0$) of the solar disk at 17 GHz, respectively, inside two bold rectangles shown in Figure 2 at the right panel.

|  | 08-Jan-2015 | 04-Dec-2016 |
|---|---|---|
| Standard deviation (σ) of V (K) | 2.3 | 5 |
| Signal threshold (5 σ) of V (K) | 11.5 | 25 |
| Average of $I_0$ (K) at 17 GHz | 10378 | 10200 |

The two black thick rectangles in Figure 2, in the right at the bottom and top panels, show the radio-quiet Sun regions used in this study. The radio-quiet regions were analyzed for the evaluate of standard deviation of circularly polarized signal, minimum detectable signal level of circular polarized and average intensity of the solar disk. Table 1 presents the estimated parameters in radio-quiet Sun regions. The standard deviations of circularly polarized signal ($\sigma_V$) observed on 08-Jan-2015 and 04-Dec-2016 were 2.3 K and 5 K, respectively, after the images were averaged over 4-minute intervals in these regions. The average intensity ($I_0$) of the solar disk on 08-Jan-2015 was about of 10378 K while on 04-Dec-2016 was found to be 10200 K at 17 GHz.

The minimum detectable signal level of Stokes V was calculated as a five-sigma (5 $\sigma$), which were about 11.5 and 25 K for both radio-quiet Sun regions, see Table 1.

**3.2. Radio Circular Polarization, Spectra and Magnetic Field**

From Figures 3(a) and 4(a) that the radio intensity map at 17 GHz is superimposed on the optical magnetograms observed with HMI at 03:00

UT on January 8, 2015 and December 4, 2016, respectively. The brightness temperatures of the right- and left-handed circular polarized component are presented in Figures 3(b) and 4(b). The red and blue contours are the positive and negative components of the radio circular polarization, respectively. The circular polarization at AR 12257 is up to 200 K and at AR 12615 is up to 600 K for the negative polarity region. The circular polarization degree at 17 GHz are overlaid on the HMI which is presented in Figures 3(c) and 4(c). The positive components are superimposed in red, and negative components are in blue contours. The circular polarization degree at ARs 12257 and 12615 are up to 2 % and 3 % in positive and negative polarity regions, respectively. Figures 3(e) and 4(e) show EUV images at 304 Å of AIA data. The black contours exhibit radio intensity at 17 GHz. Radio intensity is shown with respect to the bright region at 304 Å. Figures 3(f) and 4(f) also display EUV images at 171 Å observed by AIA. The red and blue contours represent the radio circular polarization degree at 17 GHz.

### 3.3. Coronal and Chromospheric Magnetic Fields Components Compared with the Photospheric Magnetic Fields Derived from The Extrapolated Coronal Fields

The radio magnetic fields are estimated from the observed radio intensity, radio circular polarization, and spectral index of the brightness temperature, see formula 1 for details. The spectral index is about 0.6 around the AR. Hence, the radio magnetic fields in the positive (red) and negative (blue) components are 80, 110 and 160 G at AR 12257, see Figures 3(d) and 4(d). The radio field strengths are 80, 140 and 220 G in the positive (red) and negative (blue) polarities at AR 12615. The location of radio and photospheric magnetograms are adopted over 400 x 400 pixel around the ARs. Since the location is within the beam size of the

NoRH data at 17 GHz, which is about 10″. The comparison clearly demonstrates a good agreement in the structure of the magnetic fields between the optical and radio magnetograms.

The radio and photospheric magnetic fields are compared within 40 pixel$^2$ (~20″) regions. Results of radio and photospheric magnetic fields are estimated as emission average values in the adopted regions. At the center of active region, the chromospheric and coronal components are included in the radio polarization and cannot be separated. For the estimate the magnetic fields, we adopt some footpoints positions in negative and positive polarities, as shown in Figures 3(c) and 4(c) at regions 1 and 4, respectively. For both polarities, the radio magnetic fields are 192 G and -180 G at AR 12257. The minus sign (-) refers to the negative polarity region. The radio magnetic fields of AR 12615 are 351 G and -457 G. The corresponding photospheric magnetic fields at AR 12257 are 854 G and -782 G. Ultimately, the optical magnetic fields at AR 12615 have 910 G and -783 G in two polarities. Consequently, the comparison ratio between photospheric and radio magnetic fields is 0.22 in the positive polarity and is 0.23 in the negative polarity at AR 12257. We note that the ratio is quite similar for the AR 12257. While the ratio between photospheric and radio magnetic fields in the AR 12615 is 0.34 in the positive patch and is 0.58 in the negative region.

The microwaves radio polarizations observed by NoRH at 17 and 34 GHz refer to the upper chromosphere. The NoRH observations can be used for acquiring information on the coronal magnetic fields of the Sun. The coronal magnetic fields are derived from the coronal loops of ARs. The observed radio emission is obtained from the SDO/AIA data, the photospheric magnetic fields observed by HMI are weak at these levels. The chromospheric component was neglected for the estimate the coronal magnetic fields. Zirin et al. (1991) and Grebinskij et al. (2000) developed their methods for dealing with the observed brightness temperatures of the NoRH data. Hence, they adopted a two-component model atmosphere including the corona and chromosphere. The observed brightness temperature ($T_{b,obs}$), as a function of wavelength of observations ($\lambda$), is defined ($T_{b,obs} = T_{b,chr} + T_{b,cor}$) and the observed circular polarized is defined by ($V_{obs} = V_{chr} + V_{cor}$).

For the optically thin coronal component, the electron temperature was assumed to be $10^6$ K. While the electron temperature for optically thick of chromospheric component was $10^4$ K. Therefore, considering both of observed radio intensity at 17 GHz and electron temperatures of coronal and chromospheric components as input parameters, then the ordinary and extraordinary modes of free-free emission have two different optical depths. The coronal magnetic fields are obtained based on some assumptions as follows: a) We selected some coronal loops regions in both ARs, see Figures 3(c) and 4(c) at the numbered 2, 5, 6 and 7. b) The photospheric magnetic fields, ($B_{l,pho}$), are ~ 8 G in the selected regions. c) We adopt 40 pixel$^2$ (~20″) regions, which is more than the beam size of NoRH (10″) at 17 GHz.

**Table 2.** The coronal magnetic fields estimated from EUV observations and from the extrapolated field in the corona under the force-free approximation.

|  | Region | $I_{obs}$ (K) | $I_{chr}$ (K) | $I_{cor}$ (K) | $V_{obs}$ (K) | $B_{l,pho}$ (G) | $B_{l,cor}$ (G) | $B_{l,ff}$ (G) |
|---|---|---|---|---|---|---|---|---|
| AR 12257 | 2 | 10187.2 | 8146.2 | 2041 | -85.6768 | -6.00714 | -127.603 | 65.3620 |
|  | 5 | 10022.6 | 8094.6 | 1928 | 99.4159 | 4.29685 | 156.7436 | 92.8225 |
|  | 6 | 10156 | 8906 | 1250 | 50.1668 | 8.79262 | 121.9965 | 111.456 |
|  | 7 | 9956.63 | 8335.63 | 1621 | -64.1114 | -1.51898 | -120.225 | 91.1376 |
| AR 12615 | 2 | 11103.6 | 10132.6 | 971 | -76.938 | -3.43057 | -240.859 | 35.8637 |
|  | 5 | 10225.8 | 8403.8 | 1822 | 98.0017 | 6.7059 | 163.5032 | 91.1089 |
|  | 6 | 10086.4 | 7129.4 | 2957 | 92.9248 | 6.7342 | 95.52596 | 145.850 |
|  | 7 | 11717 | 9295 | 2422 | -174.373 | -7.63188 | -218.85 | 131.858 |

Hence, the coronal emission is derived from the AIA observations. The spectral index is close to 2 for the optically thin coronal component (Miyawaki et al. 2016). Using the formula 1, the coronal magnetic fields are obtained. These results are estimated as emission average values. In Table 2, the coronal magnetic fields are computed at these regions. We find the observed radio intensity in the range 9000-11000 K. The coronal emission from AIA data is 900-2000 K. The chromospheric intensity components were obtained from subtracting observed radio intensity with the coronal emission observed by AIA data. Then, we find that this component has almost observed total radio intensity of 7000-10000 K. The circular polarization ($V_{obs}$) is in the range 50-170 K. In the end, the coronal magnetic fields are presented in the eighth column of Table 2. From the calculation, we conclude that the coronal fields are found to be between 90-240 G in regions 2, 5, 6 and 7 of two active regions.

The structure of the extrapolated magnetic field in the corona under the force-free approximation is presented in Figure 7 on HMI magnetograms for two active regions. The extrapolation code was given by Vlahos (2017). The numerical method is proposed by Alissandrakis (1981), and it reconstructs the 3D magnetic field above the photospheric boundary, where the current density (J) is parallel to magnetic field ($B_l$). The extrapolated results represent the coronal loops that connected regions with intense magnetic fields and opposite polarities. We compare coronal magnetic fields derived from radio observations and the extrapolated force-free fields. The extrapolated coronal fields are found to be in the range 30-145 G above the photospheric boundary. We note that the coronal magnetic fields obtained from the linear force-free method are smaller than the coronal fields estimated from using the AIA data.

Figure 5 shows the radio intensity at 17 and 34 GHz and circular polarization at 17 GHz along a radial cut passing through the center of active region in Figure 2 (right panels). The AR 12615 is represented by solid lines, but the AR 12257 is presented by dashed lines. The maximum total intensity at 17 GHz is about 2.5 x$10^4$ K above active regions. The circularly polarized components have weaker signals. However, the maximum radio intensity at 34 GHz is about 1.7 x $10^4$ K in active regions. In Figure 6, the photospheric magnetic fields observed with HMI at AR 12257 (dashed line) and 12615 (solid line), along line passing through the East-West direction, are shown above two islands of magnetic fields of different polarity. The result of comparison shows that magnetic fields in both cases are quite identical with indications of opposite features in magnetic polarities between two active regions. The largest magnetic fields were about 1800 G in the positive and negative polarity regions.

## 4. Summary and Conclusion

Our main results are summarized as follows:

- From our study, the maximum radio circular polarization in AR 12257 is about 180 K in the positive polarity region, but it reaches to 200 K in the negative polarity region.
- The AR 12257 has smaller observed circular polarization than the AR 12615, where the positive and negative components of AR 12615 have radio circular polarization up to 350 K.
- We note a widespread structure on the radio intensity map concentrated in the N-S direction for AR 12257, but in the case of AR 12615 which displays a different structure elongated in the E-W direction.
- The result of circular polarization degree is found to be in the range from 0.8 to 2 % in the AR 12257, while the AR 12615 reveals the polarization degree exceeding 2.9 %. However, the intensity map of AR 12615 shows a large patch of bright source than the AR 12257, where the radio emission of ARs originates in the coronal loop. This is the reason for that the AR 12615 shows larger degree of circular polarization than AR 12257. The observed circular polarization degree is higher than the estimated values by Bogod and Gelfreikh (1980) and Iwai and Shibasaki (2013), which shows a clear difference in the comparison between the previous analyses. The observed degree of circular polarization were in the range from 0.5 % to 1.7 %, which found by Iwai and Shibasaki (2013), while Bogod and Gelfreikh (1980) obtained the degree of circular polarization up to 1.5 %.
- The ratio between photospheric and radio magnetic fields of AR 12257 is 22 % in the positive polarity region and 23 % in the negative

polarity region. For AR 12615 is between 34 % and 58 % in the positive and negative polarity regions, respectively.

- The average intensity of the solar disk on January 8, 2015 is 10378 K and 10200 K on December 4, 2016. The standard deviations of circularly polarized signal on January 8, 2015 and December 4, 2016 were 2.3 K and 5 K, respectively.

- The structure of the extrapolated magnetic field in the corona under the force-free approximation is presented above the photospheric boundary. We estimated the coronal magnetic field at the edge of ARs 12257 and 12615, where the radio circular polarization was produced mainly from coronal loops, the coronal magnetic fields are calculated to be 100-240 G.

**Acknowledgements**

The authors would thank the staff of NAOJ (in Japan) for the participation of Al-Azhar University (Egypt) in the analysis of Nobeyama Radioheliograph data at the Solar Data Analysis System (SDAS) based on a Letter of Agreement signed in 2014. SDO data are courtesy of NASA/SDO and the AIA and HMI science terms. The extrapolation force-free code has been kindly provided by professor Loukas Vlahos, which is much appreciated by the authors.

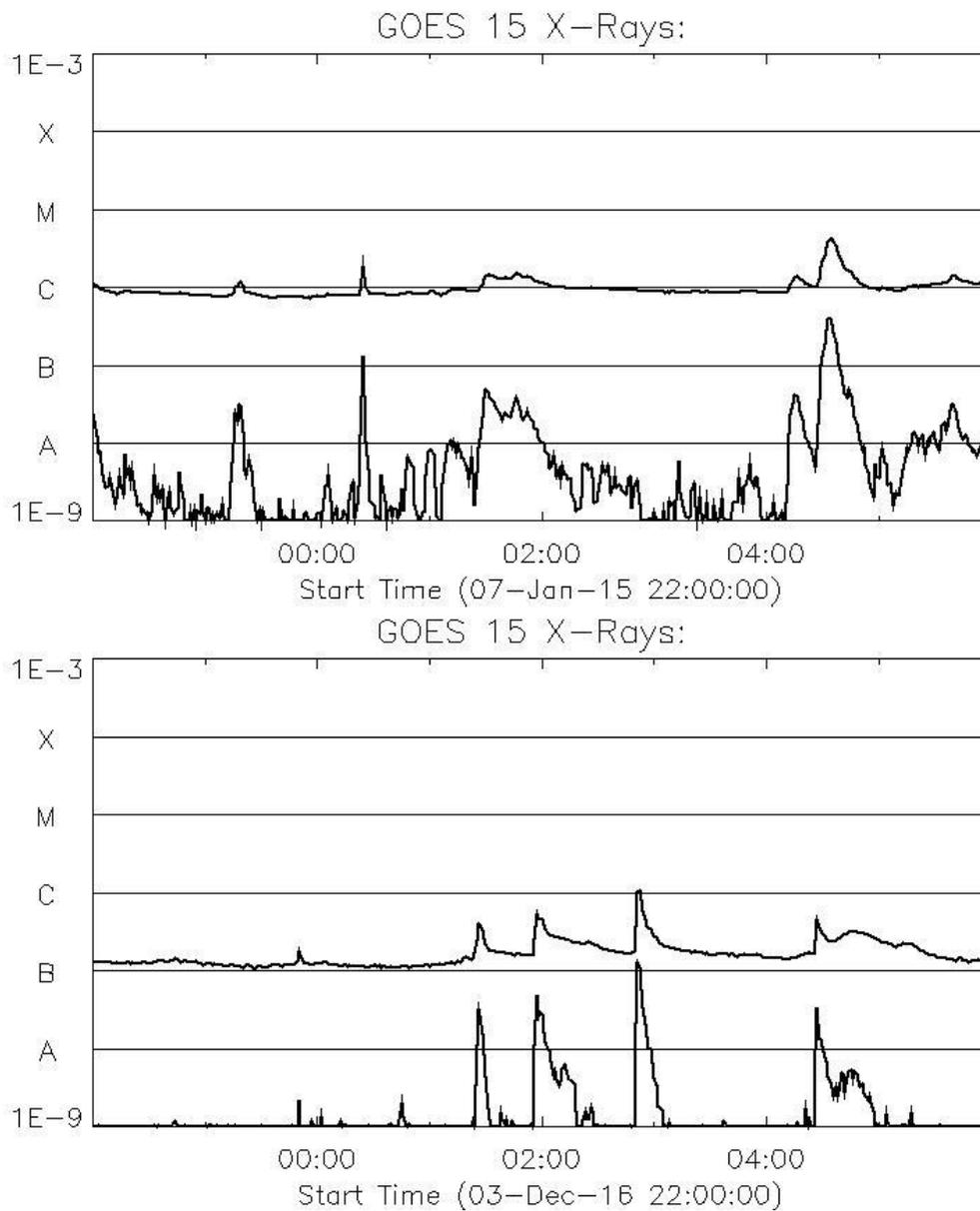

Figure 1: Time variation of the total soft X-ray flux of the Sun observed with GOES 15 of the ARs 12257 (top panel) and 12615 (bottom panel).

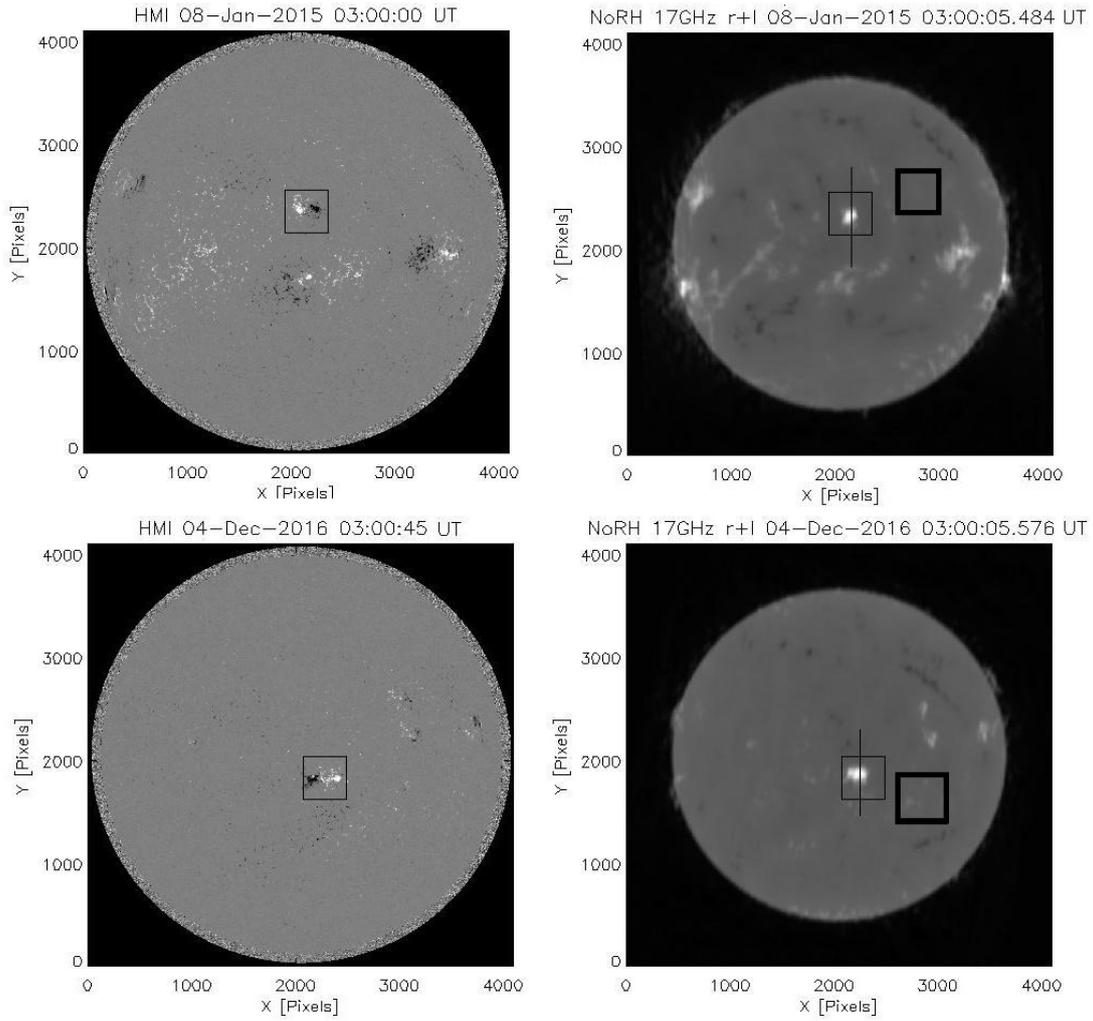

Figure 2 :( Right) radio intensity at 17 GHz observed with NoRH at 03:00:05 UT on January 08, 2015 and December 04, 2016. Thick rectangles display radio-quiet regions explained in Table 1. (Left) SDO/HMI magnetic field at ~ 03:00:00 UT. Thin rectangles found in right and left panels show two analyzed active regions.

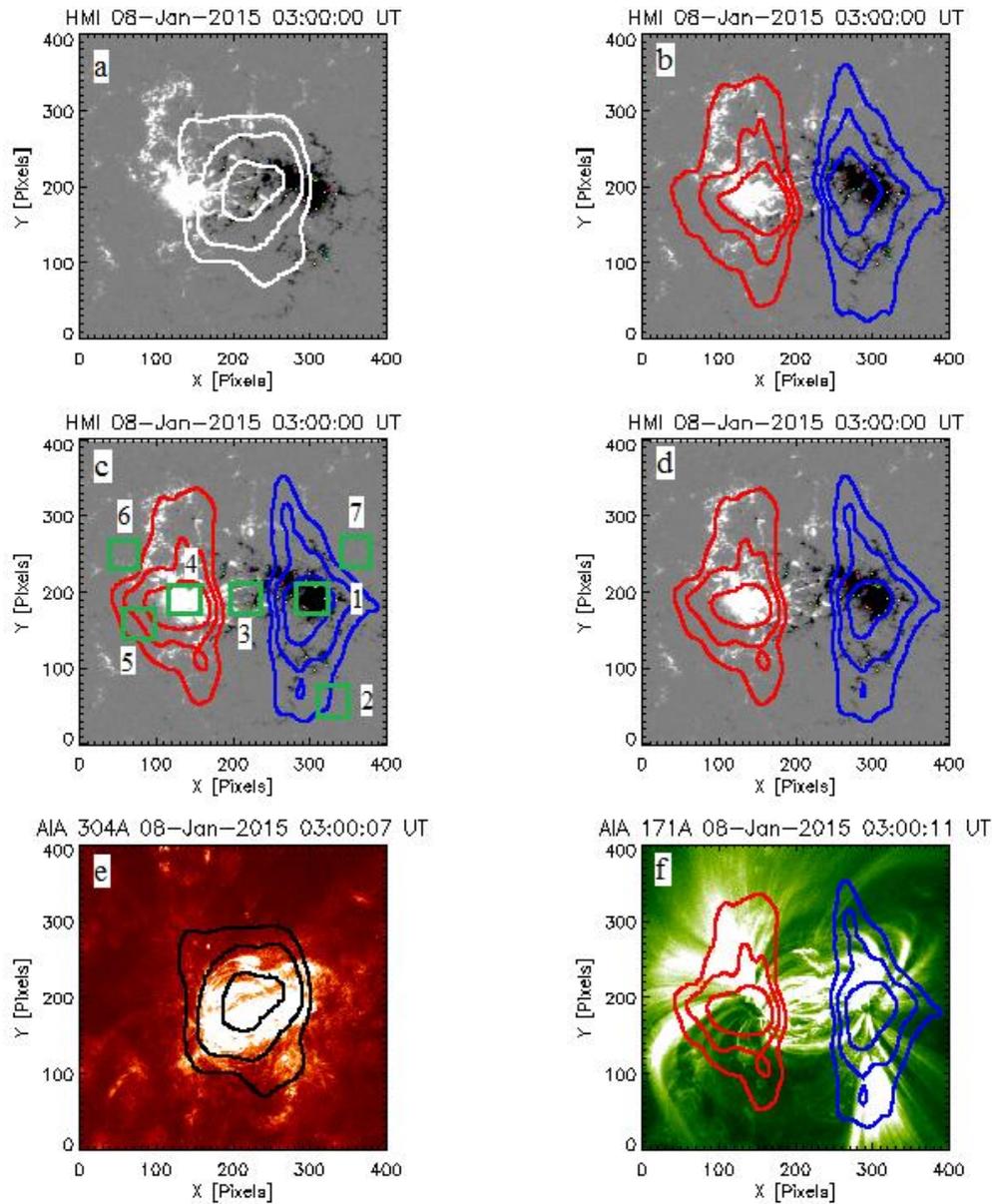

**Figure 3.** (**a**) Magnetic fields observed by HMI at 03:00 UT on January 8, 2015 and radio intensity at 17 GHz is represented by white contours. (**b**) Photospheric magnetic fields and radio circular polarization at 17 GHz are shown in red for positive components: 75, 130, 180 K and in blue for negative components: 75, 130, 200 K. (**c**) Photospheric magnetic fields and radio circular polarization degree at 17 GHz are displayed in red for the positive polarity: 0.8 %, 1.1 %, 1.5 % and in blue for the negative polarity: 0.8 %, 1.1 %, 1.5 %. (**d**) Photospheric magnetogram observed by HMI and radio magnetic fields at 17 GHz are presented in red for positive components: 80, 110, 160 G and in blue for negative components: 80, 110, 160 G. (**e**) EUV image at 304 Å observed by AIA. Black contours: radio intensity at 17 GHz. (**f**) EUV image at 171 Å taken by AIA. Red and blue contours show positive and negative radio circular polarization degree at 17 GHz, respectively. As shown in panel (c), footpoints in negative and positive polarities are shown by regions 1 and 4, respectively. The loop top is indicated in region 3. While at the edge of AR in negative and positive polarities are represented in regions 2 and 5, respectively.

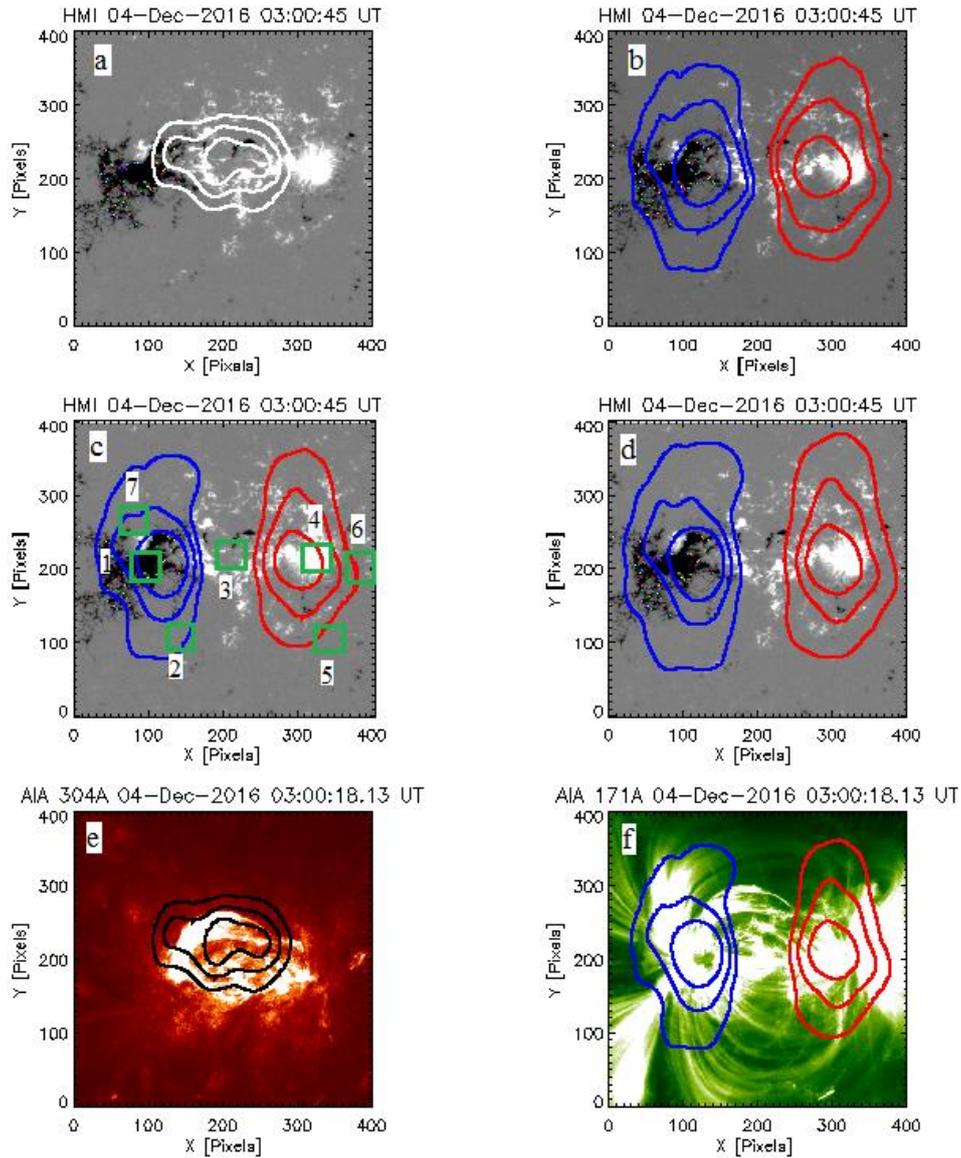

**Figure 4. (a)** Magnetic fields observed by HMI at 03:00 UT on December 4, 2016 and radio intensity at 17 GHz is represented by white contours. **(b)** Photospheric magnetic fields and radio circular polarization at 17 GHz are shown in red for positive components: 100, 160, 350 K and in blue for negative components: 100, 160, 350 K. **(c)** Photospheric magnetic fields and radio circular polarization degree at 17 GHz are displayed in red for the positive polarity: 1 %, 1.5 %, 2.5 % and in blue for the negative polarity: 1 %, 1.5 %, 2.5 %. **(d)** Photospheric magnetogram observed by HMI and radio magnetic fields at 17 GHz are presented in red for positive components: 80, 140, 220 G and in blue for negative components: 80, 140, 220 G. **(e)** EUV image at 304 Å observed by AIA. Black contours: radio intensity at 17 GHz. **(f)** EUV image at 171 Å taken by AIA. Red and blue contours show positive and negative radio circular polarization degree at 17 GHz, respectively. As shown in panel (c), footpoints in negative and positive polarities are shown by regions 1 and 4, respectively. The loop top is indicated in region 3. However, at the edge of AR in negative and positive polarities are represented in regions 2 and 5, respectively.

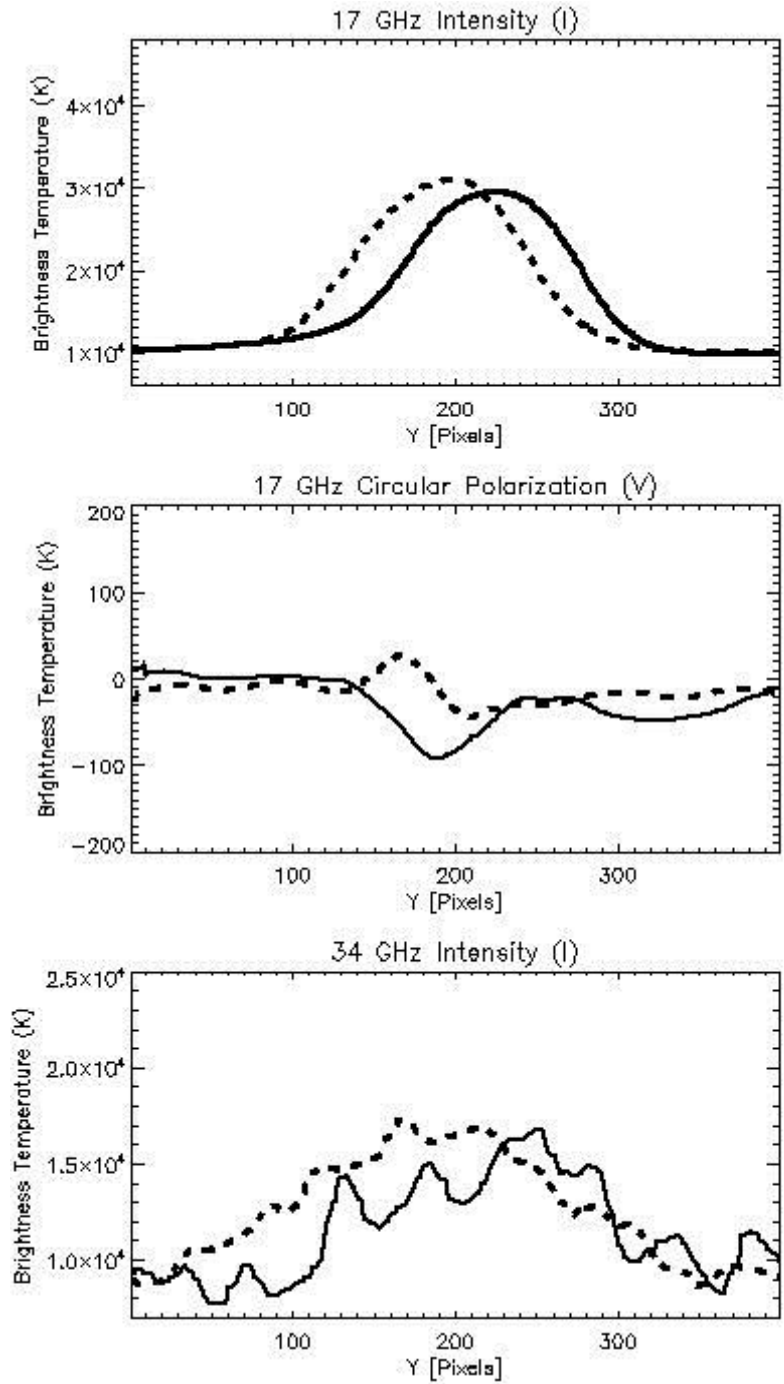

**Figure 5.** (Top) radio intensity, (middle) circular polarization at 17 GHz, and (bottom) radio intensity at 34 GHz along the black line passing through the North-South direction over the active regions, as shown in Figure 2. The AR 12615 is shown by solid lines, while the AR 12257 is presented by dashed lines.

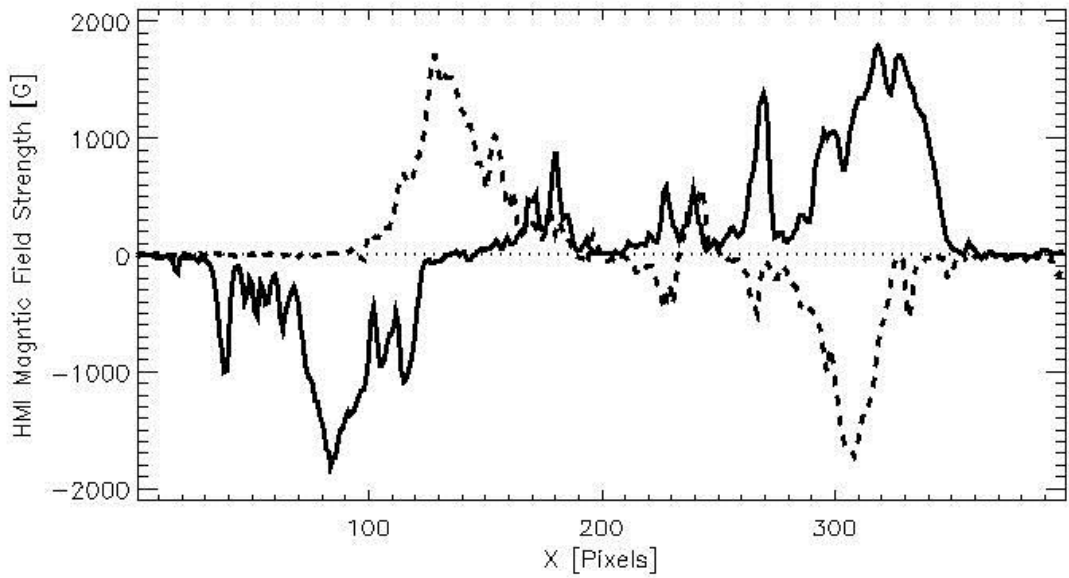

**Figure 6.** Magnetic fields observed by HMI for the ARs 12257 and 12615 along the line passing through the East-West direction over two main islands of magnetic fields of different polarity. The AR 12615 is shown as solid line, but the AR 12257 is displayed by dashed line.

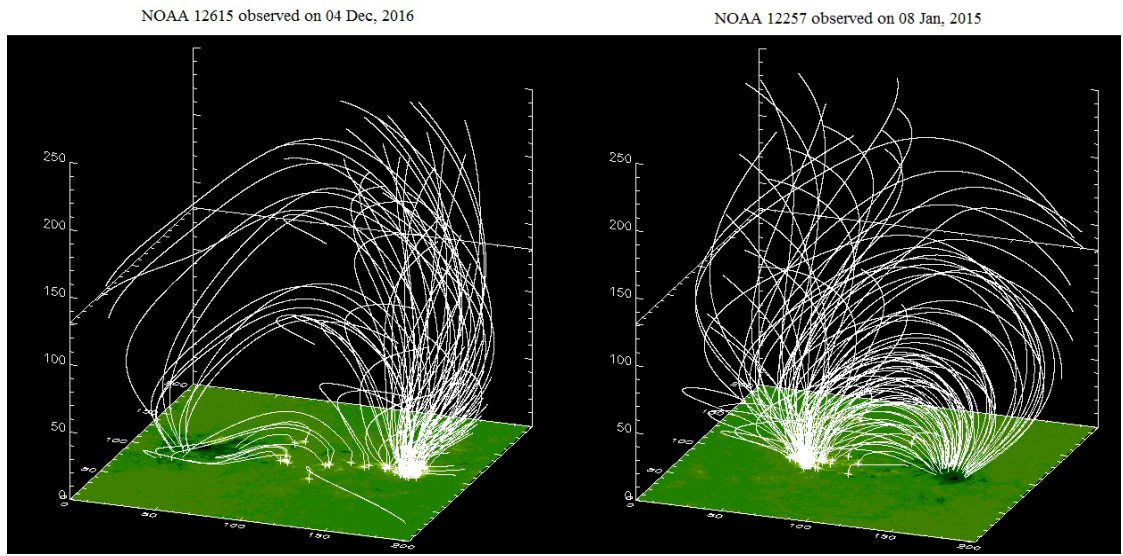

**Figure 7.** The extrapolated magnetic field lines in the ARs 12257 and 12615 under the force-free approximation.